\documentclass[final,subeqn]{siamltex}

\usepackage{amssymb}
\usepackage{graphicx}

\def\cF{{\cal F}}

\def\cL{{\cal L}}

\newcommand{\mI}{{\bf I}}
\newcommand{\mL}{{\bf L}}
\newcommand{\mA}{{\bf A}}

\newcommand{\vF}{{\bf F}}
\newcommand{\vU}{{\bf U}}

\def\tr{{\triangle r}}
\def\tth{{\triangle \theta}}

\title{Computation of Spiral Spectra 
\thanks{Mathematics Institute, University of Warwick, Coventry, CV4 7AL} }

\author{Paul Wheeler
\thanks{{\tt wheeler@maths.warwick.ac.uk}.}
\and Dwight Barkley
\thanks{{\tt barkley@maths.warwick.ac.uk}.}
}

\begin{document}

\maketitle

\begin{abstract}

A computational linear stability analysis of spiral waves in a
reaction-diffusion equation is performed on large disks.  As the disk radius
$R$ increases, eigenvalue spectra converge to the absolute spectrum predicted
by Sandstede and Scheel. The convergence rate is consistent with $1/R$, except
possibly near the edge of the spectrum.  Eigenfunctions computed on
large disks are compared with predicted exponential forms.  Away from the edge
of the absolute spectrum the agreement is excellent, while near the edge
computed eigenfunctions deviate from predictions, probably due to finite-size
effects.  In addition to eigenvalues associated with the absolute spectrum,
computations reveal point eigenvalues.  The point eigenvalues and associated
eigenfunctions responsible for both core and far-field breakup of spiral waves
are shown.

\end{abstract}



\pagestyle{myheadings}
\thispagestyle{plain}
\markboth{P. WHEELER AND D. BARKLEY}{Spiral Spectra}

\section{Introduction}
\label{sec:intro}

\begin{figure}[ht]
\begin{center}
\includegraphics[width=4cm, height=4cm]{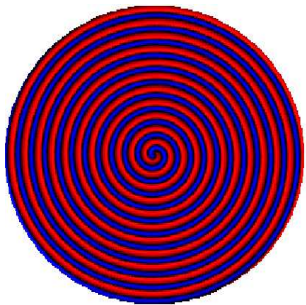}
\caption{Rotating spiral wave solution of reaction-diffusion equations
  described in \S \ref{sec:methods}.  Colors indicate the level of the $u$
  field with blue used for $u$ near zero and red used for $u$ near 1. The wave
  rotates counterclockwise.  The domain radius, $R=80$, is approximately 10
  times the spiral wavelength.  Homogeneous Neumann boundary conditions,
  corresponding to zero chemical flux, are imposed at the domain boundary.
  Model parameters are $a=0.75$, $b=0.0006$, and $\epsilon$=0.0741}
\label{fig:intro_spiral}
\end{center}
\end{figure}

Rotating spiral waves are found in many chemical and biological systems and
have been the subject of intense study for many
years~\cite{mcph93,rKkS95,jkjs98,Winfree87}.  The equations governing these
systems are typically of reaction-diffusion type.  Although each system is
modeled in detail by specific equations --- which are often very complex ---
generic features of the spiral waves can be understood from reaction-diffusion
equations with simple nonlinearities.  Figure~\ref{fig:intro_spiral} shows a
spiral wave in a generic model reaction-diffusion system described in detail
in \S \ref{sec:methods}. For the model parameters in
figure~\ref{fig:intro_spiral} the spiral wave rotates with constant frequency
and shape, i.e.\ it is a rotating wave.
  
\begin{figure}[ht]
\begin{center}
\includegraphics[width=3cm]{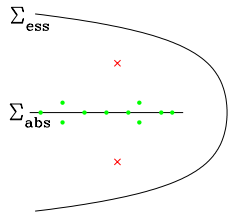}
\caption{Illustration of spectra in the complex plane for spirals on bounded
  and unbounded domains.  $\Sigma_{abs}$ and $\Sigma_{ess}$ represent the
  absolute and essential spectrum respectively. Points represent eigenvalues
  on a large bounded domain which approach $\Sigma_{abs}$ as the domain size
  tends to infinity.  Crosses represent the point spectrum which does not
  approach $\Sigma_{abs}$ as the domain size tends to infinity. }
\label{fig:intro_sketch}
\end{center}
\end{figure}

The focus of our work is a computational study of the linear stability spectra
of rotating spiral waves such as those shown in figure~\ref{fig:intro_spiral}.
To explain the motivation behind this study it is necessary to first recall
the recent analysis by Sandstede and Scheel~\cite{bsas00b, bsas00, bsas01,
bsasup} on the spectra of rotating spiral waves.  Their work examines spectra
on large bounded disks and on unbounded domains. The results can be summarized
as follows (see figure \ref{fig:intro_sketch}).  On large bounded disks, the
linear stability spectrum consists of point eigenvalues and what is called the
absolute spectrum.  The absolute spectrum is not actually part of the
stability spectrum.  However, all but possibly a finite number of point
eigenvalues converge to the absolute spectrum as the domain size tends to
infinity.  That is, except for finitely many eigenvalues that are created
through the underlying pattern as a whole, or possibly by the boundary
conditions, all eigenvalues on large bounded domains are expected to be close
to the absolute spectrum. The point eigenvalues have well-defined limits as
the domain size tends to infinity.

In practice the absolute spectrum must be computed numerically for any given
reaction-diffusion equation, e.g.~\cite{bsas00}.  Such computations require
discretization in only one space dimension and thus are relatively simple
compared with computing eigenvalues of the full stability problem on a large
domain, such as in figure ~\ref{fig:intro_spiral}.

For spiral waves on the unbounded plane, the linear stability spectrum
consists of point eigenvalues and the essential spectrum. The essential
spectrum is continuous spectrum and is determined only by the far-field waves
trains of the spiral. It too is relatively easy to compute numerically in one
space dimension. The point eigenvalues again depend on the underlying spiral
pattern as a whole.

\begin{figure}[ht]
\begin{center}
\includegraphics[width=8cm]{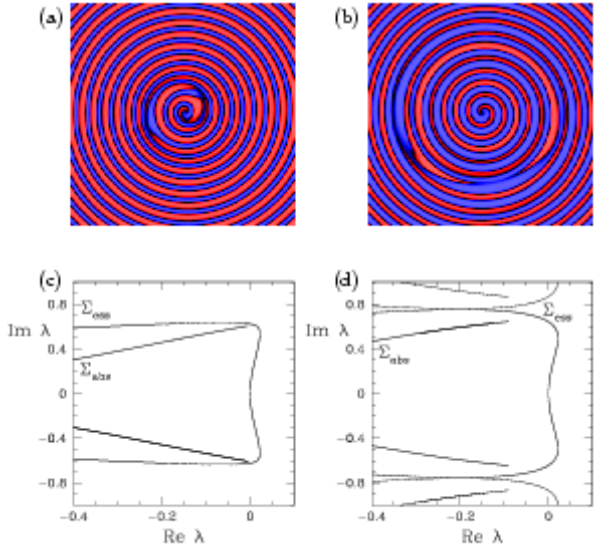}
\caption{Two examples of spiral breakup -- core breakup on the left and
  far-field breakup on the right.  The top plots show the $u$ chemical field
  at about the time of breakup in numerical simulations in square geometries
  with homogeneous Neumann boundary conditions. Domains are of length 160 on a
  side.  The bottom shows the absolute and essential spectrum obtained by
  Sandstede and Scheel for the parameter values used in the simulations.  Note
  that these spectra repeat periodically in the imaginary direction, but this
  can only be seen in (d).  Model parameters are: (left) $a=0.75$, $b=0.0006$,
  and $\epsilon=0.0741$, (right) $a=0.84$, $b=-0.045$, and $\epsilon=0.075$.}
\label{fig:breakup}
\end{center}
\end{figure}

To see how these linear stability spectra may be relevant in practice, we show
in figure~\ref{fig:breakup} simulations of two instabilities of rotating waves
on relatively large domains and the corresponding absolute and essential
spectra obtained by Sandstede and Scheel~\cite{bsas00}.  In each case a
rotating wave becomes unstable in a rather dramatic fashion and the spiral
breaks up.  Multiple spiral waves appear in each of these simulations shortly
after the time shown.  In figure~\ref{fig:breakup}(a) the breakup initiates in
the central region of the spiral and is referred to as core
breakup~\cite{mbme93,aK93,bsas00} whereas in figure~\ref{fig:breakup}(b) the
breakup first takes place in the outer regions of the spiral and is called
far-field breakup~\cite{mbmo98,qojf96,bsas00,smTeK98,lzqo00}.

The case of far-field breakup, figure~\ref{fig:breakup} (right), has been the
subject of several past
studies~\cite{ialalkaw92,mblb04,mbmo98,lbatmb03,qojf96,bsas00,stmpek98,smTeK98,
lzqo00}.  The breakup can be mostly understood from analysis and
simulations of one-dimensional systems.  While many of these studies are based
on the complex Ginzburg-Landau equation, results appear to be similar for the
case of reaction-diffusion equations~\cite{mblb04,mbmo98,smTeK98}.  The typical
scenario is that as a parameter is varied the spiral first becomes
convectively unstable.  In a bounded domain the onset of convective
instability does not generally lead to breakup because unstable modes
typically propagate away from the core and are not reflected at the boundary.
As the parameter is varied further the spiral becomes absolutely
unstable. Only at the absolute-instability threshold will instability surely
occur in a bounded domain.  Absolute instability corresponds to a growing
``global mode''~\cite{stmpek98,smTeK98}, which here means an eigenfunction on
the bounded domain whose eigenvalue has positive real part.

In the analysis of Sandstede and Scheel, convective instability is signaled by
the crossing of the essential spectrum into the right half
plane~\cite{bsas00b,bsas00}.  Figure~\ref{fig:breakup}(d) shows that the
spiral is convectively unstable.  However, the spiral was already convectively
unstable prior to the breakup seen in figure~\ref{fig:breakup} and this is not
the cause of breakup.  The breakup is caused by an eigenvalue with positive
real part and the corresponding global mode on the finite (bounded) domain.
In principle such an eigenvalue could be associated with the absolute spectrum
or it could be a point eigenvalue.  In figure~\ref{fig:breakup}(d) the
absolute spectrum is away from the imaginary axis and thus is not expected to
play a direct role in the far-field breakup. Thus we expect there to be at
least one positive point eigenvalue not contained in
figure~\ref{fig:breakup}(d).

It is worth being clear about potentially confusing terminology.  Absolute
instability is not associated only with the absolute spectrum.  The union of
the absolute spectrum and the point spectrum determine absolute stability.  If
part of the absolute spectrum lies in the right half plane, then the spiral
will necessarily be absolutely unstable.  However, the converse is not true,
since absolute instability can arise due to point eigenvalues, even if the
absolute spectrum lies entirely in the left-half plane.

We should warn the reader that simulations at the stated parameters in
figure~\ref{fig:breakup}(b) are very sensitive to numerical resolution.  As we
shall see, this is because the particular parameters considered by Sandstede
and Scheel are extremely close to a transition between far-field and core
breakup.

The case of core breakup, figure \ref{fig:breakup}(left), has not been
extensively analyzed, in large part because one cannot expect to capture much
of the spiral core structure in one-dimensional studies. (See however
\cite{mblb04}.)  Sandstede and Scheel~\cite{bsas00} have raised the
possibility that core breakup may be due to the absolute spectrum.  In figure
\ref{fig:breakup}(c) it can be seen that the absolute spectrum is very near
the imaginary axis, although it is entirely in the left-half plane.  (The
argument of Sandstede and Scheel is not simply that the absolute spectrum is
close to the real axis, but the details are not important here.)  The other
possibility for core breakup is that the instability is again due to point
eigenvalues.  The essential spectrum in \ref{fig:breakup}(c) extends into the
right-half plane and so the spiral is convectively unstable.

Computing the eigenvalue spectra of spiral waves on large domains has thus
become important.  First and foremost, it is desirable to test the absolute
spectra of Sandstede and Scheel in at least a few cases.  The primary issue is
whether or not eigenvalues tend to the absolute spectrum for typical domains
sizes used in studies of spiral waves, e.g.\ domains such as those in
figure~\ref{fig:breakup}. The theory is still developing and we would like to
know whether absolute spectra in fact have any implications for domains of
reasonable size.  The other important issue which computations can address is
the abundance and importance of point eigenvalues not predicted by the
absolute spectrum.  For example, it is desirable to know how many point
eigenvalues are present within the region shown in \ref{fig:breakup}(c)-(d),
how many of these eigenvalues have positive real part, and whether or not
these are associated with breakup.  For these reasons we have undertaken the
large-scale eigenvalue computation reported here.

Throughout this paper we shall use {\em point eigenvalue} to mean those
eigenvalues which remain isolated as the domain radius becomes large, as
contrasted with the eigenvalues associated with the absolute spectrum that
approach one another as the radius becomes large. To simplify discussion we
shall use {\em positive eigenvalue} to mean an eigenvalue, or a
complex-conjugate pair of eigenvalues, with positive real part.  Similarly we
shall use {\em positive eigenfunction} to mean an eigenfunction whose
corresponding eigenvalue has positive real part.

\section{Model and Methods}
\label{sec:methods}

\subsection{Model}

We will consider a standard two-component reaction-diffusion
model~\cite{Barkley91} given by the equations
\begin{subequations}\label{eq:rd1}
\begin{eqnarray}
\frac{\partial u}{\partial t} & = & 
\nabla^2u + f(u,v) \\ 
\frac{\partial v}{\partial t} & = & \delta \nabla^2v + g(u,v),
\end{eqnarray}
\end{subequations}
where
\begin{equation}
f(u,v) = \frac{1}{\epsilon}u(1-u)\left(u-\frac{v+b}{a}\right)
\end{equation}
There is freedom in the choice of $g(u,v)$ and our methods will not depend on
this choice. However, the results we report will be for the $g$ proposed by
B\"ar and Eiswirth~\cite{mbme93} and used by Sandstede and
Scheel~\cite{bsas00}, namely
\begin{equation}
g(u,v) = \left\{
\begin{array}{ll}
-v, & 0 \leq u < 1/3 \\
1-6.75u(u-1)^2 - v, & 1/3 \leq u \leq 1 \\
1-v, & 1 < u
\end{array}\right.
\end{equation}

The equations are posed on a disk of radius $R$ and with homogeneous Neumann
boundary conditions at $r=R$:
$$
\frac{\partial u}{\partial r}(R,\theta) =
\frac{\partial v}{\partial r}(R,\theta) = 0,
$$
where $r$, $\theta$ are standard polar coordinates.  For chemically reacting
systems these are the most natural boundary conditions as they correspond to
zero chemical flux through the boundary of the domain.  Other boundary
conditions could give different spiral solutions and linear stability spectra
on finite domains, but we do not consider any other boundary conditions here.

The parameters of the model are kinetics parameters $a$, $b$, and $\epsilon$,
and the diffusion coefficient $\delta$.  If $b>0$ the equations model an
excitable medium. In this case the homogeneous state with $u=v=0$ everywhere
is linearly stable and finite amplitude perturbations are required to initiate
waves. The perturbation threshold is set by $b/a$.  For $b<0$ the equations
model an oscillatory medium. In both cases $\epsilon$ controls the time-scale
ratio between the $u$- and $v$-equations.  We consider $\epsilon \ll 1$
corresponding to a fast time scale for $u$ relative to $v$.  We shall only
report results for the case $\delta=0$. $\delta=1$ is the other case commonly
considered.  As stated at the outset, these equations model generic features
of spiral waves in a variety of excitable and oscillatory media.

\subsection{Computational tasks}

Consider rotating-wave solutions of (\ref{eq:rd1}) rotating at frequency
$\omega$.  We use $(u^*, v^*)$ to denote such solutions and refer to them as
steady spirals, since these are steady states when viewed in the frame of
reference which is rotating with the spiral.  Transforming to a system of
coordinates co-rotating at frequency $\omega$, steady spirals obey the
equations
\begin{subequations}\label{eq:rd2}
\begin{eqnarray}
0 & = & \nabla^2 u^* 
+ \omega \frac{\partial u^*}{\partial \theta} + f(u^*,v^*) \\
0 & = & 
\omega \frac{\partial v^*}{\partial \theta} + g(u^*,v^*),
\end{eqnarray}
\end{subequations}
subject to homogeneous Neumann boundary conditions.
These steady-state equations can be written in the form
\begin{equation}
\label{eq:ss}
\cF 
\left(\begin{array}{c} u^*\\ v^* \end{array}\right)
= 0,
\end{equation}
where $\cF$ is the nonlinear operator given by the right hand side of
(\ref{eq:rd2}).

Next, given a steady spiral, we seek the leading part of its linear stability
spectrum.  Consider the linearized evolution equations, in the rotating frame,
for infinitesimal perturbations $(u^\prime,v^\prime)$ of the steady solution
$(u^*,v^*)$:
\begin{subequations}\label{eq:pert}
\begin{eqnarray}
\frac{\partial u^\prime}{\partial t} & = & \nabla^2 u^\prime 
+ \omega \frac{\partial u^\prime}{\partial \theta} 
+ f_u(u^*,v^*)u^\prime + f_v(u^*,v^*)v^\prime \\
\frac{\partial v^\prime}{\partial t} & = & 
\omega \frac{\partial v^\prime}{\partial \theta} 
+ g_u(u^*,v^*)u^\prime + g_v(u^*,v^*)v^\prime,
\end{eqnarray}
\end{subequations}
where $f_{u}, \dots, g_v$ denote the derivatives of the kinetic functions.
In this frame of reference the eigenvalue problem is
\begin{equation}
\label{eq:eval}
\cL \left(\begin{array}{c} \tilde u\\ \tilde v \end{array}\right)
= \lambda \left(\begin{array}{c} \tilde u\\ \tilde v \end{array}\right)
\end{equation}
where $\left(\begin{array}{c} \tilde u\\ \tilde v \end{array}\right)$ are
eigenfunctions, $\lambda$ are the corresponding eigenvalues, and $\cL$ is
\begin{equation}
\label{eq:cL}
\cL = \left(\begin{array}{cc} \nabla^2 + \omega\partial_\theta + f_{u}(u^*,
v^*) & f_{v}(u^*, v^*) \\ g_{u}(u^*, v^*) & \omega\partial_\theta +
g_{v}(u^*, v^*)
\end{array}\right).
\end{equation}
Thus our primary numerical tasks are the solution of steady state problem
(\ref{eq:ss}) and the determination of the leading eigenvalues of problem
(\ref{eq:eval}).

In addition it is necessary to perform a few simulations of the time-dependent
equations~(\ref{eq:rd1}), e.g., the simulations shown in
figure~\ref{fig:breakup}.  In the case of figure~\ref{fig:breakup} the
numerical methods are described fully elsewhere~\cite{Barkley91,DowleM97} and
will not be discussed here.

\subsection{Computational Methods}

Equations (\ref{eq:ss}) and (\ref{eq:eval}) are common in large-scale
numerical bifurcation problems and the computational methods we employ are
more or less standard~\cite{NumBif}.  For completeness we provide a basic
description of our methods and stress a few points concerning implementation
which are essential to the efficiency of the computations.

The fields are discretized on a regular polar grid $(r_j, \theta_k) = (j \tr,
k \tth)$, where $0 < j \leq N_r$ and $0 \leq k < N_\theta$, plus the center
point $(0,0)$.  There are thus $N_r N_\theta +1$ grid points.  The
$r$-derivatives in the differential operators are evaluated using second-order
finite differences, taking into account the boundary condition at $r=R$.  The
$\theta$-derivatives are evaluated spectrally using Fourier transforms. In
this way (\ref{eq:ss}) and (\ref{eq:eval}) become
\begin{eqnarray}
\label{eq:ss_num}
\vF(\vU^*) = 0, \\
\label{eq:eval_num}
\mL \tilde \vU = \lambda \tilde \vU 
\end{eqnarray}
where the $\vU$'s are vectors of length $N = 2(N_r N_\theta +1)$, $\vF$ is a
nonlinear function, and $\mL$ is an $N \times N$ matrix. 

Newton's method is used to solve steady state problem (\ref{eq:ss_num}).  One
iteration of Newton's method is
\begin{eqnarray}
\label{eq:netwon}
{\bf DF}(\vU_n) \triangle \vU_n = -\vF(\vU_n)  \\
\vU_{n+1} = \vU_{n} + \triangle \vU_n,
\end{eqnarray}
where ${\bf DF}(\vU_n)$ is the linearization of $\vF$ about the current
iterate $\vU_n$. This is the same matrix as $\mL$ except it is evaluated at
$\vU_n$ rather than at the steady state $\vU^*$.

\begin{figure}[ht]
\begin{center}
\includegraphics[width=3.5cm]{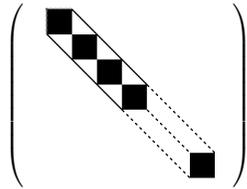}
\caption{Banded structure of matrix ${\bf DF}$ or $\mL$. Solid black
represents full blocks of size $2 N_\theta \times 2 N_\theta$ mainly due to
using spectral representation of $\theta$-derivatives. The lines are due to
the second-order finite-difference representation of $r$-derivatives.}
\label{fig:matrix}
\end{center}
\end{figure}

The work of each Newton's iteration is dominated by the work necessary to
solve the $N \times N$ linear system of equations (\ref{eq:netwon}) for the
$n^{th}$ correction $\triangle \vU_n$.  This can be done by a direct method if
care is taken to order variables to keep the matrix bandwidth of ${\bf DF}$ as
small as possible.  Let $u_{jk}$ and $v_{jk}$ be values at grid point $(r_j,
\theta_k)$. Then these are ordered in $U_i$ such that the chemical species
changes fastest with index $i$, followed by the angular index $k$, followed by
the radial index $j$.  This ordering is not that suggested by (\ref{eq:cL}).
With the ordering we use the bandwidth is approximately $4 N_\theta$.  See
figure~\ref{fig:matrix}. Even for moderately large discretizations a direct
method can be used to solve (\ref{eq:netwon}).  For example on a grid $N_r
\times N_\theta = 600 \times 256$, $N$ is approximately $3 \times 10^5$, while
the bandwidth is only about $10^3$.

The only other issue concerning the steady state computations is that the
frequency $\omega$ must be found in addition to fields $u^*$ and $v^*$.  The
existence of the additional unknown is consistent with the fact that the
solution to (\ref{eq:rd2}) is not unique due to the rotational symmetry in
$\theta$. One more algebraic equation must be added to (\ref{eq:ss_num}) to
remove the phase degeneracy and thus giving $N+1$ equations in $N+1$ unknowns.
The constraint we add is simply to fix $u$ as some point in the domain.  While
the constraint destroys the banded structure of ${\bf DF}$, a Sherman-Morrison
technique~\cite{NRinC} is used to find solutions of the augmented linear
system using only the banded ${\bf DF}$.

\begin{figure}[ht]
\begin{center}
\includegraphics[width=10cm]{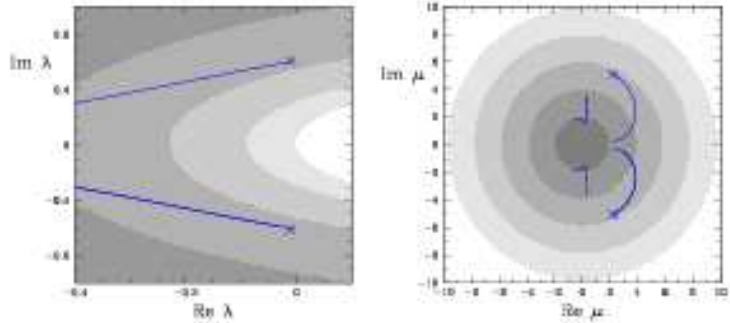}

\caption{Plot showing the effect of the Cayley transform on the absolute
  spectrum in \ref{fig:breakup}(c).  Left is the original and right is
  transformed spectrum.  The right includes two periodic repeats of the
  absolute spectrum which are outside of the region shown on the left.  
  Shading indicates magnitude of eigenvalues after the transformed.}
\label{fig:cayley}
\end{center}
\end{figure}

We now describe our computations of the leading eigenvalue spectrum of $\mL$.
The basis of our approach is to employ a Cayley transformation to transform
the eigenvalues we seek (those with largest real part) to dominant eigenvalues
(of largest magnitude), and then to find iteratively dominant eigenvalues of
the transformed operator. Reference~\cite{Meerbergen96} gives a nice review of
such methods.  Specifically, we consider the matrix $\mA$ defined by the
Cayley transform
\begin{equation}
\label{eq:cayley}
\mA = (\xi\mI + \mL)^{-1} (\eta\mI + \mL),
\end{equation}
where $\xi$ and $\eta$ are real parameters and $\mI$ is the identity.  Letting
the $\mu$ and $\lambda$ be the eigenvalues of $\mA$ and $\mL$ respectively, we
have the relation
\begin{equation}
\label{eq:cayley_ev}
\mu = \frac{\eta + \lambda}{\xi + \lambda}.
\end{equation}
The parameters $\xi$ and $\eta$ can be adjusted so as to map the regions of
interest in the $\lambda$-plane to large magnitude in the $\mu$-plane.  Using
the predicted absolute spectra of Sandstede and Scheel it is easy to find
appropriate values of $\xi$ and $\eta$.  Figure \ref{fig:cayley} shows the
effect of the Cayley transform on the absolute spectrum for one of the cases
predicted by Sandstede and Scheel~\cite{bsas00} (the other case is similar)
for the values of $\xi$ and $\eta$ used in our computations: $\xi=-0.4$ and
$\eta = 4.0$.  In the $\mu$-plane we include transforms of two periodic
repeats of the absolute spectrum, one corresponding to the larger imaginary
part and one to smaller imaginary part (in the $\lambda$-plane ).  These
repeats are outside the region of the $\lambda$-plane shown.  Most of the
eigenvalues of $\mL$ lie far to the left in the $\lambda$-plane (outside the
range of the figure). These are all mapped to near the origin by
(\ref{eq:cayley}).

There is no need to form or store the matrix $\mA$ in order to iteratively
calculate its eigenvalues. All one needs is the ability to compute $\mA \vU$
for arbitrary $\vU$.  This is accomplished using the same basic technique as
in Newton's method.  Letting $\vU' = \mA \vU$ we see that $\vU'$ obeys
\begin{equation}
\label{eq:cayley_num}
(\xi \mI + \mL) \vU' = (\eta \mI + \mL) \vU.
\end{equation}
However, $(\xi\mI + \mL)$ has the same structure, in particularly the same
bandwidth, as $\mL$ and $(\eta\mI + \mL)$ requires mostly the same
computations as evaluating $\vF$.  Thus we act with $\mA$ on $\vU$ by
computing $(\eta\mI + \mL)\vU$ to form a right-hand side and then solving a
linear system with matrix $(\xi\mI + \mL)$.  Since this is a fixed matrix, for
any given $\mL$, it is factored only once for all actions of $\mA$.

Dominant eigenvalues $\mA$ are easily found by subspace
iteration~\cite{Meerbergen96,TrefethenNLA}.  This method guarantees that we
can obtain any required number of dominant eigenvalues to arbitrarily high
precision.  While Arnoldi's method generally converges faster, in practice we
find that with this method all eigenvalues we require are not found with high
enough precision. While there are methods, such as block Arnoldi, which could
probably address this, we have used subspace iteration.  From the eigenvalues
$\mu$ of $\mA$ we invert (\ref{eq:cayley_ev}) to find the required eigenvalues
$\lambda$.

\subsection{Accuracy}
\label{sec:accuracy}

We conclude this section by considering the accuracy of our computations and 
providing details of numerical parameters used in the results reported.
The sources of error are the following:
\begin{enumerate}
\item
Discretization error of the steady state problem, i.e.\ approximation of
(\ref{eq:ss}) by (\ref{eq:ss_num}).

\item
Residual error arising from determining the roots of (\ref{eq:ss_num})
to a finite accuracy.

\item
Discretization error of the eigenvalue problem, i.e.\ approximation of
(\ref{eq:eval}) by (\ref{eq:eval_num}).

\item
Residual error arising from computing eigenvalue/eigenfunction pairs
(\ref{eq:eval_num}) to finite accuracy.
\end{enumerate}

\begin{figure}[ht]
\begin{center}
\includegraphics[width=8cm]{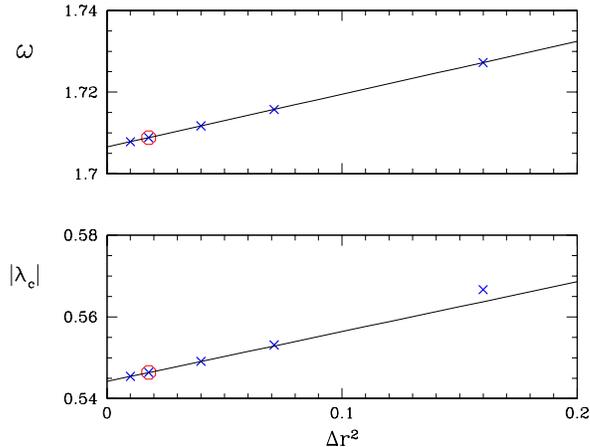}

\caption{Graphs illustrating the convergence of steady states (top) and
  eigenvalues (bottom) as function of grid resolution $\triangle r^2$, where
  $\omega$ is the spiral frequency and $\lambda_c$ is a complex eigenvalue.
  The domain radius is $R=40$.  Crosses are with $N_\theta=256$.  For
  $\triangle r = 0.1333$ computations have been repeated with $N_\theta=128$
  are results are shown with circles.  Parameters are $a=0.75$, $b=0.0006$,
  and $\epsilon=0.0741$. }
\label{fig:conv}
\end{center}
\end{figure}

The two residual errors are least important.  We always iterate sufficiently
to reduce these to negligible size.  The following hold for all reported
results.  Solutions $\vU^*$ of (\ref{eq:ss_num}) are found such that
$\|\vF(\vU^*)\| < 10^{-8}$. Solutions of (\ref{eq:eval_num}) are found such
that $\|\mL \tilde \vU - \lambda \tilde \vU \| < 10^{-8}$, where $\|\tilde \vU
\| = 1$.  The norm is the vector 2-norm.  The dimension $k$ of the subspaces
used in subspace iteration are: $k=30$ for $R=20$ and $R=40$, and $k=75$ for
$R=80$.  In all cases we stop iterations when $\sim 0.7 k$ of the
eigenvalue-eigenvector pairs have residual less than $10^{-8}$.  In the case
of $R=80$, we thus obtain 53 pairs with the required residual.  We initially
start with a subspace generated from $k$ random vectors, but we restart
iterations from previous runs when necessary.

The discretization errors are dominated, in both the steady state and
eigenvalue computations, by the second-order finite-derivative approximation
to the $r$-derivatives in the Laplacian operator. This is expected since the
$\theta$-derivatives are computed with spectral accuracy.  Thus the dominant
error in the computations depends on $\triangle r$ in a well-understood way.
Figure \ref{fig:conv} shows examples of how results from steady state and
eigenvalue computations scale with $\triangle r^2$.  The domain radius is
$R=40$, half the maximum considered in our work.  For the steady states we
show the frequency $\omega$ and for the eigenvalues we show the magnitude of
$\lambda_c$, a complex eigenvalue associated with core breakup that will be
considered in detail later in the paper.  In both cases the second-order
convergence is evident.  We are interested only in leading eigenvalues
(roughly $10^2$ out of $10^5$) all of which correspond to eigenfunctions with
variation on approximately the same spatial scale (roughly the wavelength of
the underlying spiral) so that the effects of the finite-difference
discretization will be approximately the same for all eigenvalues we report.

Based on these plots, we use $\triangle r = 0.1333$ for all results reported
in \S~\ref{sec:results}.  At this value of $\triangle r$ we have performed
computations at both $N_\theta=128$ and $N_\theta=256$.  These results show
clearly that $N_\theta=128$ gives sufficient resolution for domain radius
$R=40$.  Therefore, for radii up to at least $R=80$, $N_\theta=256$ should
produce smaller errors than the already small errors due to the radial
discretization.  In summary, for all results in \S~\ref{sec:results} we use
$\triangle r=0.1333$ and $N_\theta=256$.

\section{Results}
\label{sec:results}

\subsection{Spectra}
\label{sec:spectra}

\begin{figure}[ht]
\begin{center}
\includegraphics[width=7.5cm]{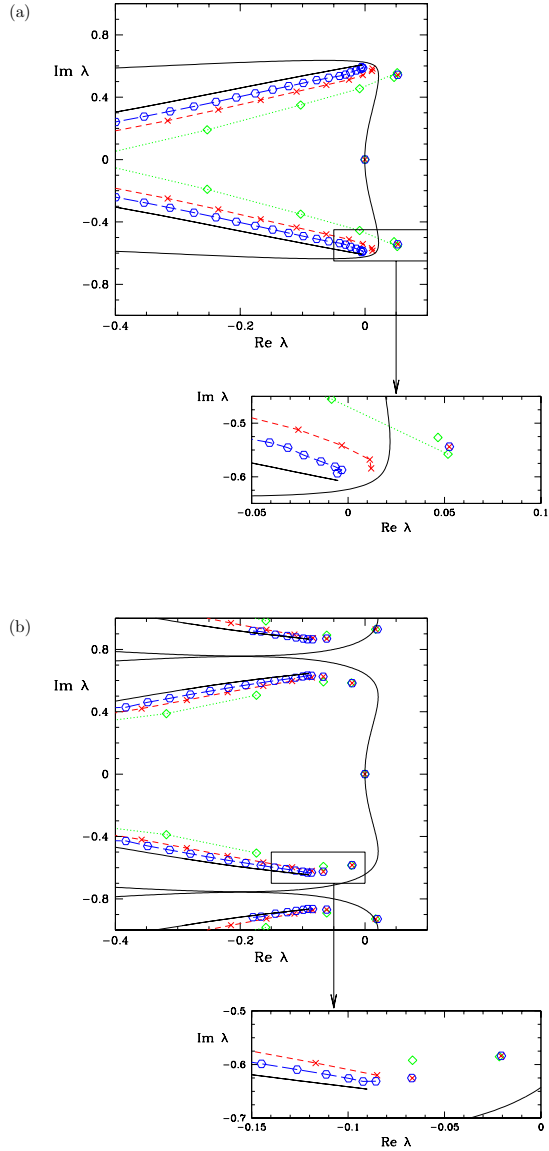}
\caption{Eigenvalue spectra. (a) Spectrum corresponding to core breakup and
(b) spectrum corresponding to far-field breakup.  In each case eigenvalues are
shown for three domain radii: $R=20$ (green diamonds), $R=40$ (red crosses),
and $R=80$ (blue circles).  For each radius, the eigenvalues associated with
the absolute spectrum are connected with lines.  Point eigenvalue are not.
Parameters for (a): a=0.75, b=0.0006, $\epsilon$=0.0741,
$\omega$=1.71. Parameters for (b): a=0.84, b=-0.045, $\epsilon$=0.0751,
$\omega$=1.50.}
\label{fig:eval}
\end{center}
\end{figure}

We begin with results for the eigenvalue spectra.  Figure
\ref{fig:eval} shows leading eigenvalues of $\mL$ for the two cases considered
by Sandstede and Scheel.  The spectrum corresponding to core breakup is at the
top and the spectrum corresponding to far-field breakup is at the bottom.  In
each figure eigenvalues computed for three domain radii, $R=20$, $R=40$, and
$R=80$, are plotted as points with dashed lines connecting eigenvalues
associated with the absolute spectrum.  For comparison, the absolute and
essential spectra obtained by Sandstede and Scheel are shown as solid curves.

Before considering either case in detail, we note that the predominant feature
in both is the presence of many eigenvalues which approach the predicted
absolute spectra as the domain radius increases.  Each doubling of $R$ results
in approximately a halving of the distance of eigenvalues to the absolute
spectrum, thus supporting $1/R$ convergence to the absolute spectrum.
Furthermore the density of eigenvalues approximately doubles with each
doubling of $R$.  We return to this while considering each case in more
detail.  It should be noted that in the far-field case we have not obtained
all eigenvalues associated with the periodic repeats of the absolute spectrum
for $R=80$ due to the difficulties of computing these with sufficient
accuracy.  This results in an abrupt termination of the eigenvalue branches
for $R=80$ at the top and bottom of figure~\ref{fig:eval}(b).

Consider first the spectrum in figure~\ref{fig:eval}(a) corresponding to core
breakup.  Within the region of the complex plane shown, there are three point
eigenvalues.  All other eigenvalues are associated with the absolute spectrum.
Specifically, we find three eigenvalues which are insensitive to the domain
radius and which are separated from the absolute spectrum.  Of these, one is
the zero eigenvalue arising due to rotational symmetry.  There are three
points indistinguishable from zero in figure~\ref{fig:eval}(a) corresponding
to the three domain radii studied.  The other two point eigenvalues are a
complex-conjugate pair at approximately $0.050 \pm 0.543 i$.  As we shall show
in \S~\ref{sec:breakup}, these eigenvalues are associated with core
breakup. We will denote them by $\lambda_c$.  (These are the eigenvalues
considered in the convergence study in figure~\ref{fig:conv}.)  All other
eigenvalues vary with domain radius and approach the absolute spectrum as the
radius becomes large.

The enlargement in figure~\ref{fig:eval}(a) clarifies the situation around the
complex point eigenvalues.  Even on the enlarged scale the point eigenvalues
for $R=40$ and $R=80$ coincide.  At $R=20$, however, there are two nearby
eigenvalues. The lower one corresponds to the absolute spectrum (indicated by
the connecting dashed line) because it moves, as $R$ is increased, toward the
absolute spectrum. The other eigenvalue converges, as the $R$ is increased, to
the point eigenvalue.  Note that while the edge of the eigenvalue branch
associated with the absolute spectrum is near the point eigenvalue at $R=20$,
the branch does not approach the point eigenvalue as the domain becomes small.
It is nevertheless interesting that the point eigenvalue is near the edge of
the absolute spectrum. We find this throughout and return to this in the
conclusion.

As already noted, the distance of eigenvalues to the absolute spectrum is
approximately proportional to $1/R$ and the density of eigenvalues is
approximately proportional to $R$.  Because we are not able to extend the
computations significantly beyond the radius $R=80$ (already quite large)
there is not enough data to draw strong conclusions about the these scalings.
In particular it is not clear from the data whether or not the scaling in the
vicinity of the edge of the absolute spectrum is different from elsewhere.
Near the edge of the spectrum the eigenvalues are more dense and closer to the
absolute spectrum than elsewhere.  More importantly we observe a curving and
perhaps folding, at the edge of the eigenvalue branch as the radius becomes
large. This would again suggest a different scaling at the spectrum's edge, but
the numerical results are inconclusive.

We have focused our study on the eigenvalues within the region shown in
figure~\ref{fig:eval}(a), but we have computed some eigenvalues out side of
this region.  In particular our iterative technique frequently finds
eigenvalues associated with the periodic repetition of the absolute spectrum
in the complex plane.  We have not attempted to resolve the details of the
other eigenvalue branches.  Also there is a complex-conjugate pair of point
eigenvalues near $0 \pm i \omega$ due to approximate translational symmetry.

Now consider the spectrum corresponding to far-field breakup.  In
figure~\ref{fig:eval}(b) we find four complex-conjugate eigenvalue pairs that
we can clearly classify as point eigenvalues. One of these pairs has small 
positive real part and hence the spiral wave is absolutely linearly unstable,
see section \S~\ref{sec:breakup}.  Again we observe that the point
eigenvalues, except for the zero eigenvalue, appear near the edge of the
absolute spectrum.

We observe approximately $1/R$ convergence of eigenvalues to the absolute
spectrum. The only apparent deviation is again at the edge of the spectrum.
In this case we do not observe any curving of the eigenvalue branch seen in
the enlargement in figure~\ref{fig:eval}(a); however, we do find that the
right-most point of the computed branch does not appear to approach the edge
of the predicted absolute spectrum. One possibility is that this last
eigenvalue is in fact a point eigenvalue very close to the edge of the
absolute spectrum.

\subsection{Eigenfunctions}

\begin{figure}[h]
\begin{center}
\includegraphics[width=12cm]{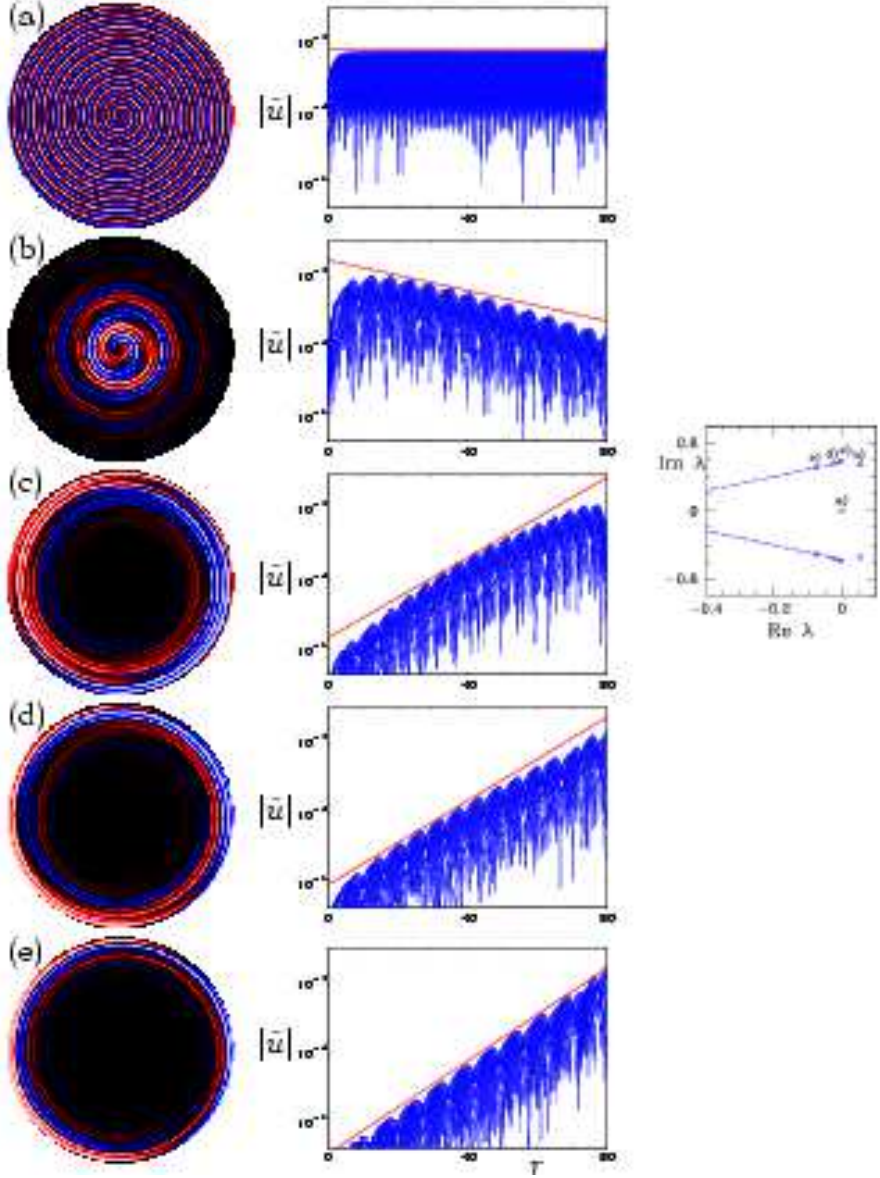}
\caption{Representative eigenfunctions for parameters corresponding to core
  breakup.  The left column shows the real part of the each eigenfunction. The
  $\tilde{u}$-field is plotted with black used where the field is near zero.
  Right column shows the $r$ dependence of $|\tilde{u}|$ with predicted
  growth/decay rate also shown with lines (see text).  Plot at the far right
  is a guide to the corresponding eigenvalues.  (a) zero (rotational)
  eigenvalue, (b) positive eigenvalue $\lambda_c$, and (c), (d), (e) three
  representative eigenvalues associated with the absolute spectrum.
  Parameters as in figure~\ref{fig:eval}(a).}
\label{fig:s1evec}
\end{center}
\end{figure}

\begin{figure}[h]
\begin{center}
\includegraphics[width=12cm]{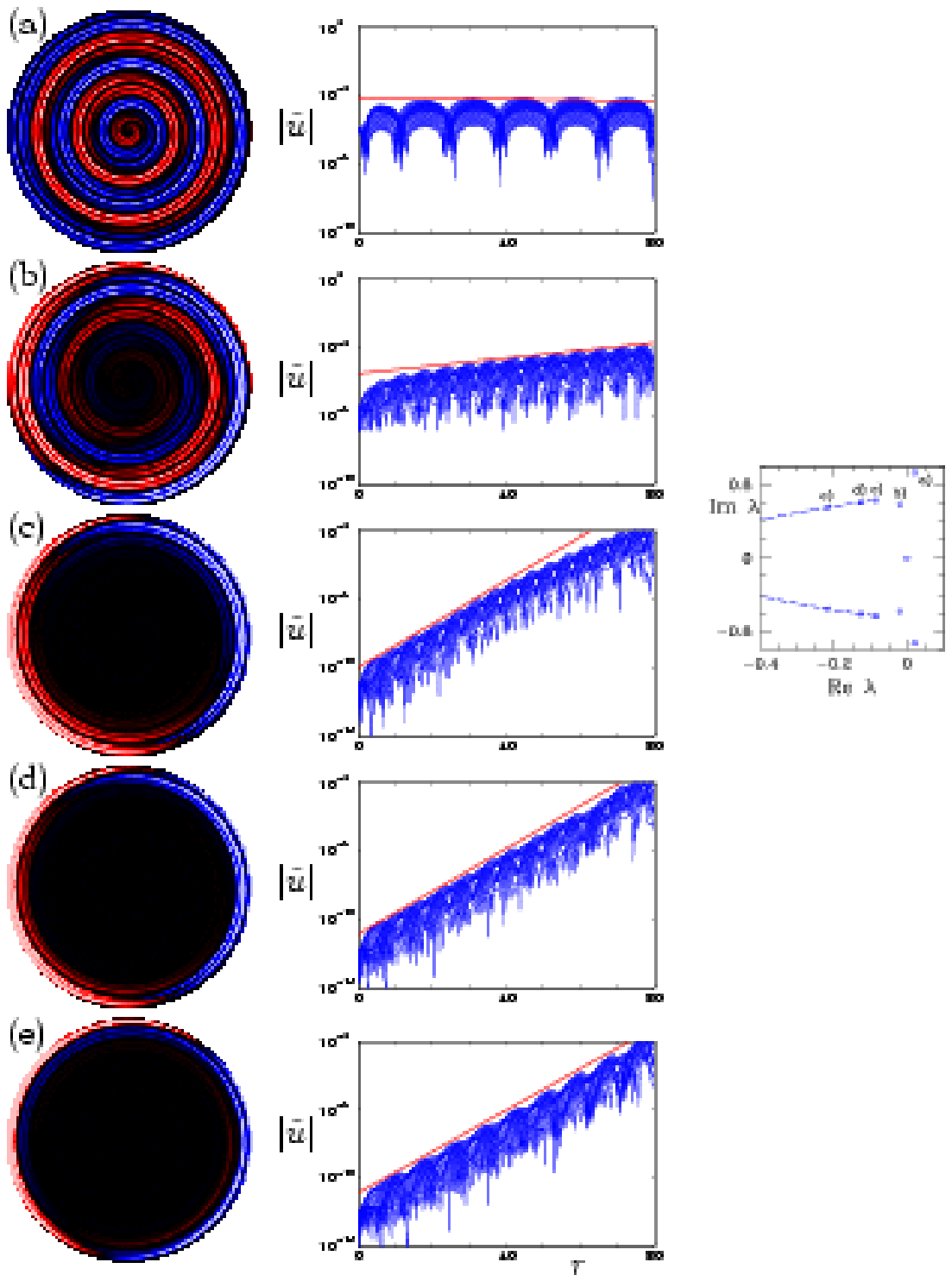}
\caption{Representative eigenfunctions for parameters corresponding to
  far-field breakup. Same format is used as in figure~\ref{fig:s1evec}.  The
  rotational eigenfunction is not shown. Two eigenfunctions corresponding to
  point eigenvalues are shown with (a) slightly positive eigenvalue and (b)
  slightly negative eigenvalue.  (c), (d), (e) are three representative
  eigenvalues associated with the absolute spectrum.  Parameters as in
  figure~\ref{fig:eval}(b).}
\label{fig:s2evec}
\end{center}
\end{figure}

In figures~\ref{fig:s1evec} and \ref{fig:s2evec} we plot eigenfunctions for a
representative selection of eigenvalues.  All eigenfunctions have been
obtained on a domain with $R=80$, the largest we consider.  Each eigenfunction
is shown in two formats.  In the left column eigenfunctions are visualized on
the computational domain. Specifically the $\tilde{u}$-field of the real part
of the eigenfunction is plotted with black used where the field values are
near zero.  The imaginary parts of the eigenfunctions and not fundamentally
different.  In the right column the modulus of each eigenfunction is shown as
a function of radius.  Specifically, we generate $16$ radial sections
$|\tilde{u}(r,\theta_s)|$ where $\theta_s = s \pi/8$ and $s = 0, 1, \dots, 15$
and plot all $16$ sections simultaneously as a function of $r$.  The envelope
of these sections gives a simple representation of the modulus of the
eigenfunction as a function of $r$.

The technique used by Sandstede and Scheel~\cite{bsas01} to obtain absolute
and essential spectra also predict large-$r$ behavior of eigenfunctions. The
main prediction is that if an eigenvalue is to the left of the essential
spectrum then the corresponding eigenfunctions will be exponentially growing
at large $r$, whereas if an eigenvalue is to the right of the essential
spectrum then the corresponding eigenfunctions will be exponentially decaying
at large $r$.  In addition to the general prediction, the numerical technique
employed by Sandstede and Scheel to obtain spectra for specific problems also
provides the growth/decay rates for eigenfunctions.  These
rates~\cite{bsemail} are indicated by the (red) lines in the right column of
figures~\ref{fig:s1evec} and \ref{fig:s2evec}.  For eigenfunctions
corresponding to the absolute spectrum, the predicted exponential growth rates
have been taken from the point on the absolute spectrum nearest to the
computed eigenvalue.  Only the slope of the lines is relevant. The intercept
is chosen for ease of comparison with the eigenfunctions.

Consider first the eigenfunctions in figure~\ref{fig:s1evec} corresponding to
the case of core breakup. The top eigenfunction is the zero mode due to
rotational symmetry. This eigenfunction is given by the $\theta$-derivative of
the underlying spiral wave and hence closely resembles the spiral. The
eigenfunction neither grows nor decays at large $r$.

Figure~\ref{fig:s1evec}(b) shows the eigenfunction corresponding to the
positive complex-conjugate point eigenvalues $\lambda_c$.  Since the
eigenvalues are to the right of the essential spectrum the eigenfunction
decays with $r$.  While there is generally good agreement between the computed
eigenfunction and prediction, there is some deviation from prediction that is
more pronounced at larger $r$.

Figures~\ref{fig:s1evec}(c)-(e) show three eigenfunctions associated with the
absolute spectrum.  The agreement between the computed eigenfunctions and
predictions is extremely good away from the edge of the absolute spectrum.
Near the edge the agreement is less good. In particular, eigenfunctions are
not pure exponential, even at large $r$, and the computed eigenfunctions
systematically grow more slowly than prediction.  While not shown, we find
that the eigenfunctions computed on smaller domains and show even slower growth
as a function of $r$. This would suggest that the deviation shown in
figure~\ref{fig:s1evec}(c) is due to finite domain size.

Figure~\ref{fig:s2evec} shows eigenfunctions corresponding to parameters for
which far-field breakup occurs.  We plot eigenfunctions corresponding to two
of the complex-conjugate point eigenvalues and show three representative
eigenfunction associated with the absolute spectrum.  The eigenfunction
corresponding to the zero eigenvalue is similar to figure~\ref{fig:s1evec}(a)
and is not shown.

The eigenfunctions corresponding to the point eigenvalues agree very well with
prediction. The growth rate of the positive eigenfunction in
figure~\ref{fig:s2evec}(a) is very small since the corresponding eigenvalue is
almost exactly on the essential spectrum (figure~\ref{fig:eval}(b)).  This is
a fortuitous situation which illustrates nicely that the essential spectrum
delimits the crossover from growth to decay of eigenfunctions.  While
quantitatively the agreement is very good, there is a qualitative disagreement
between the computed eigenfunction and prediction.  Our computed eigenfunction
is growing with $r$, indicating that the eigenvalue is actually slightly to
the left of the essential spectrum, whereas in figure~\ref{fig:eval}(b) the
eigenvalue it slightly to the right of the essential spectrum and the
predicted exponent is slightly negative. The quantitatively the difference is
very small and is likely due to a small numerical difference, e.g.\ a
difference in the value of $\omega$ found in our computations and that used by
Sandstede and Scheel.  The closeness of this eigenvalue to the essential
spectrum is just by chance. If parameters are changed the eigenvalue moves
away from the essential spectrum. It is because of this closeness to the
eigenvalue to essential spectrum that numerical simulations at these parameter
values are so sensitive to numerical resolution (as noted in
\S~\ref{sec:intro}). 

The eigenfunction in figure~\ref{fig:s2evec}(b) is exponentially growing since
the corresponding eigenvalue is to the left of the essential spectrum. There
are no observable deviations from pure exponential growth at large $r$ and the
agreement with prediction is very good. 

The three eigenfunctions associated with the absolute spectrum show the same
trend as in figure~\ref{fig:s1evec}.  The agreement between the computed
eigenfunctions and predictions is extremely good away from the edge of the
absolute spectrum while near the edge eigenfunctions are not pure exponential
and systematically grow more slowly than prediction.  This case is even more
striking than figure~\ref{fig:s1evec} for the following reasons.  The growth
rates in figure~\ref{fig:s1evec} are much larger than figure~\ref{fig:s2evec}.
(Note the scale change.)  The numerical values representing the eigenfunctions
span a larger range and yet the computed eigenfunctions away from the edge
agree very well with predictions. There is every reason to believe that these
eigenfunctions are numerically well resolved within the finite domain.  Unlike
the case in figure~\ref{fig:s1evec}(c), here the eigenfunction closest to the
edge of the spectrum, figure~\ref{fig:s2evec}(c), deviates from exponential
growth only at large $r$. There is a clear range $r$, up to approximately
$r=40$, where the eigenfunction agrees well the predicted exponential growth.
This strongly suggests that the lack of agreement is due to finite-size
effects. It is nevertheless interesting that these are more pronounced near
the edge of the spectrum.

\subsection{Implications for breakup}
\label{sec:breakup}

\begin{figure}[ht]
\begin{center}
\includegraphics[width=7cm]{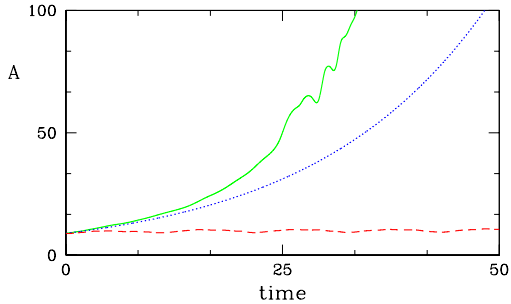}
\caption{Two time series showing the evolution starting from different
  perturbations of the steady spiral with parameters leading to core breakup.
  In one case (solid green) the perturbation is the eigenfunction
  corresponding to positive eigenvalue $\lambda_c$ [(b) in
  figure~\ref{fig:s1evec}].  In the other (dashed red) the perturbation is the
  eigenfunction corresponding to right-most eigenvalue associated to the
  absolute spectrum [(c) in figure~\ref{fig:s1evec}].  The dotted blue curve
  shows exponential growth at rate given by $\lambda_c$.  Parameters
  are as in figure~\ref{fig:eval}(a). }
\label{fig:core_perturb_ts}
\end{center}
\end{figure}

\begin{figure}[ht]
\begin{center}
\includegraphics[width=9cm]{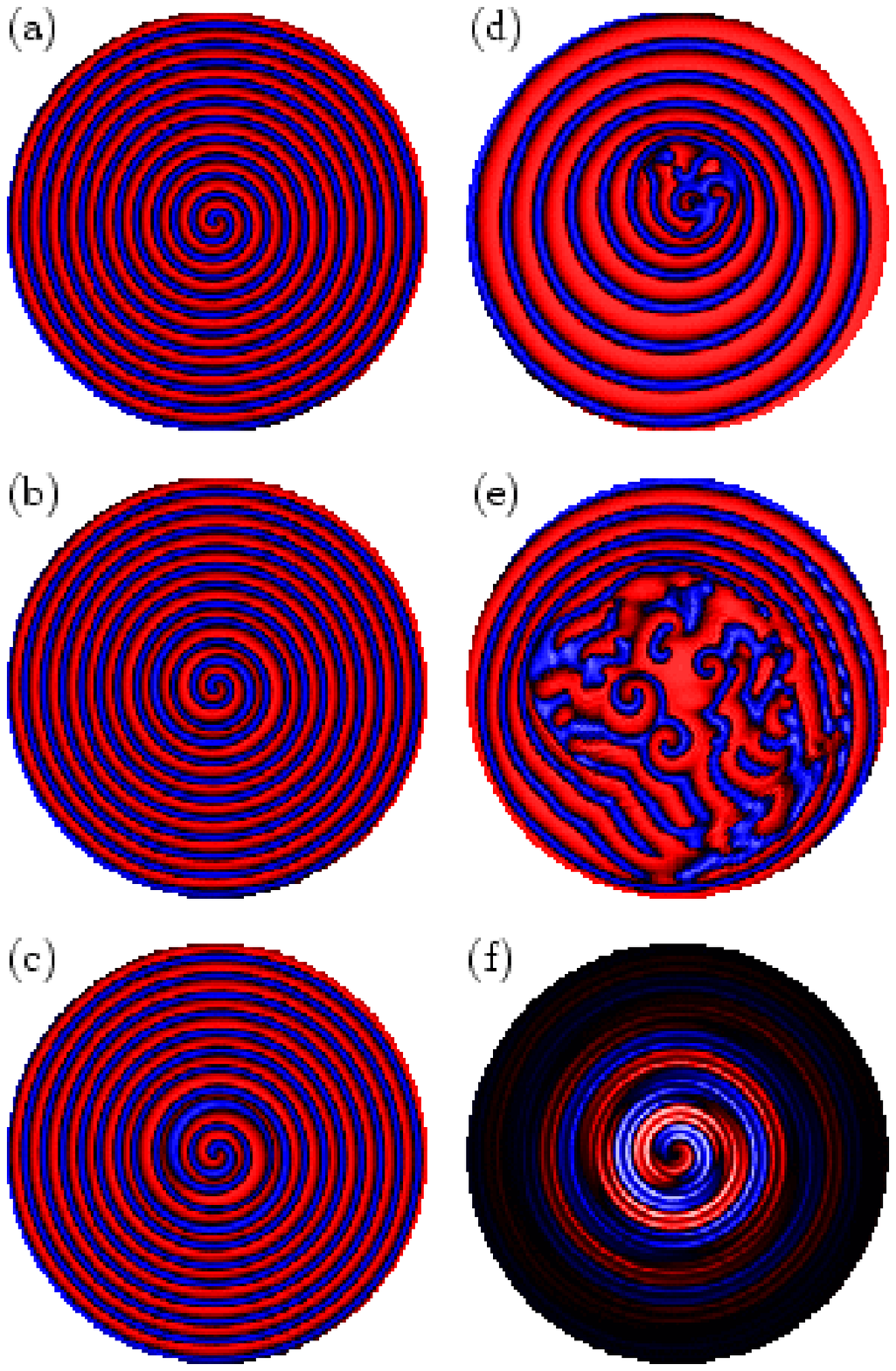}
\caption{Snapshots of evolution from the perturbed steady spiral leading to
  far-field breakup (solid green curve in figure~\ref{fig:core_perturb_ts}).
  The $u$-field is shown with $u\simeq 0$ blue and $u\simeq 1$ red.  (a)
  $t=0$, (b) $t=20$ ($\simeq 5$ periods), (c) $t=25$, (d) $t=85$ ($\simeq 23$
  periods), (e) t=120 ($\simeq 33$ periods), (f) eigenfunction.  Parameters
  are as in figure~\ref{fig:eval}(a). }
\label{fig:core_perturb_vis}
\end{center}
\end{figure}

We now return to the issue of spiral breakup discussed at the outset
(figure~\ref{fig:breakup}).  We begin with the case of core breakup.  Recall
that while this was treated by Sandstede and Scheel~\cite{bsas00,bsas01}, they
were not able to draw definite conclusions about the role of the absolute and
essential spectrum in core breakup.  It is already clear from the spectra in
figure~\ref{fig:eval} that the steady spiral is linearly unstable due to the
presence of positive point eigenvalues $\lambda_c$. 
Here we present additional nonlinear simulations of the breakup.

Figure~\ref{fig:core_perturb_ts} shows the time evolution from two different
initial conditions each composed of the steady spiral plus a small amount of
one of the computed eigenfunctions.  The amplitude plotted is defined as $A =
\min_{\, \theta} \| U^* - R_\theta U \|$, where $R_\theta$ is a rotation by
angle $\theta$. The norm is the vector 2-norm.  Essentially $A$ is the norm of
the difference between the $u$-field of the steady spiral, $U^*$, and the
$u$-field of the nonlinear solution $U$. The minimization over rotation is
included to take into account the small drift of the nonlinear solution
relative to the steady spiral.

Consider the evolution starting from the initial condition formed from the
positive eigenfunction corresponding to $\lambda_c$.  Accompanying
visualizations are presented in figure~\ref{fig:core_perturb_vis}.  The
dynamics is initially linear, obeying the exponential growth dictated by the
real part of $\lambda_c$.  After a short time the growth becomes nonlinear and
almost immediately core breakup occurs [figure~\ref{fig:core_perturb_vis}(c);
time 25]. Beyond this time the amplitude $A$ loses most of its meaning.
Visualizations at much later times are shown.  One of the more striking
aspects of the breakup is that it occurs at $r \simeq 20$, not at the center
of the spiral.  This radius is near where the unstable eigenfunction has
maximal magnitude.  Visually one sees the similarity between the nonlinear
breakup and the unstable eigenfunction in figure~\ref{fig:core_perturb_vis}.

The initial nonlinear growth in figure~\ref{fig:core_perturb_ts} is faster
than linear. This means that, at lowest order, the effect of nonlinearity on
the instability is destabilizing.  Such behavior occurs, for example,
sufficiently close to a subcritical bifurcation, e.g.~\cite{Henderson96}.
This nonlinear destabilization is consistent with the fact that small positive
eigenvalues lead to the dramatic breakup of the spiral wave. If nonlinearity
were stabilizing, one would expect the linear instability to saturate in a
state resembling a superposition of the original spiral and a small amount of
the unstable eigenmode (as occurs for example in spiral meandering
\cite{Barkley92,Barkley94,Barkley95}).  We note that in systems such as this
one the behavior can change very rapidly with parameters following a
bifurcation~\cite{stmpek98,smTeK98}, and hence we are not able to conclude
that the nonlinear growth follows from a subcritical bifurcation, only that at
these parameter values it is destabilizing.

For completeness we have also computed the nonlinear evolution from an initial
condition formed from the eigenfunction corresponding to the right-most
eigenvalue of the absolute spectrum [(c) in figure~\ref{fig:s1evec}].
Figure~\ref{fig:core_perturb_ts} shows the initial dynamics from this
simulation.  Not surprisingly $A$ does not change much on the scale of
figure~\ref{fig:core_perturb_ts} because the associated eigenvalue is very
near zero.  The simulation eventually leads to core breakup if run long
enough. However, this is simply because the steady spiral is linearly
unstable. When breakup does eventually occur, it follows the same route (e.g.\
same exponential growth) as for the initial condition based on the positive
eigenfunction.

The conclusion is that, in this case, core breakup is due to nonlinear effects
following from linear instability due to a complex-conjugate pair of point
eigenvalues.  The absolute spectrum plays no direct role in the spiral
breakup.  

\begin{figure}[ht]
\begin{center}
\includegraphics[width=7cm]{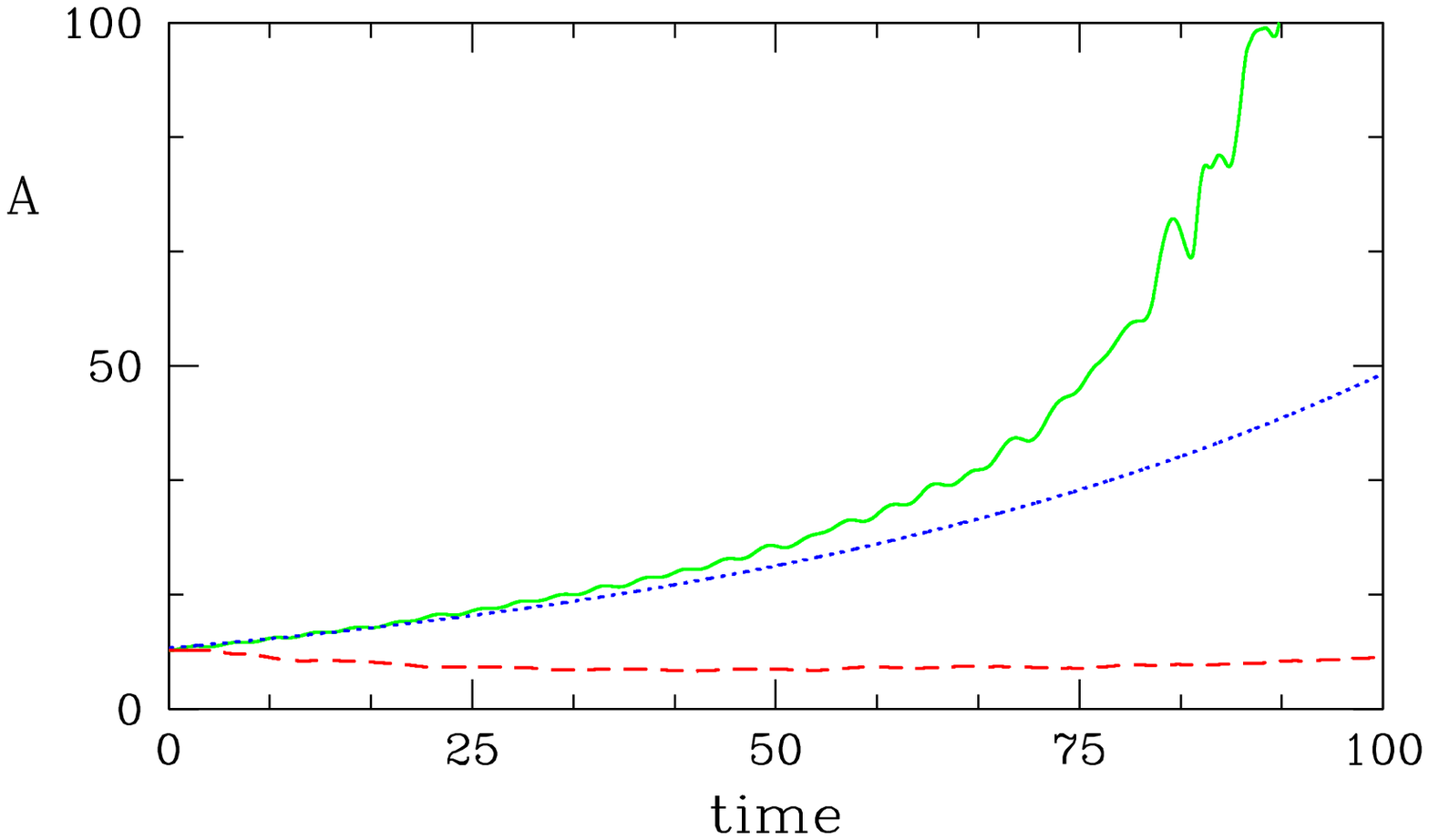}
\caption{Two time series showing the evolution starting from different
  perturbations of the steady spiral with parameters leading to far-field
  breakup.  In one case (solid green) the perturbation is the eigenfunction
  corresponding to positive eigenvalue [(a) in figure~\ref{fig:s2evec}].  In
  the other (dashed red) the perturbation is the eigenfunction corresponding
  to the weakly stable point eigenvalue [(b) in figure~\ref{fig:s2evec}].  The
  dotted blue curve shows exponential growth at rate given by the positive
  eigenvalue.  Parameters are as in figure~\ref{fig:eval}(b). }
\label{fig:far_perturb_ts}
\end{center}
\end{figure}

\begin{figure}[ht]
\begin{center}
\includegraphics[width=9cm]{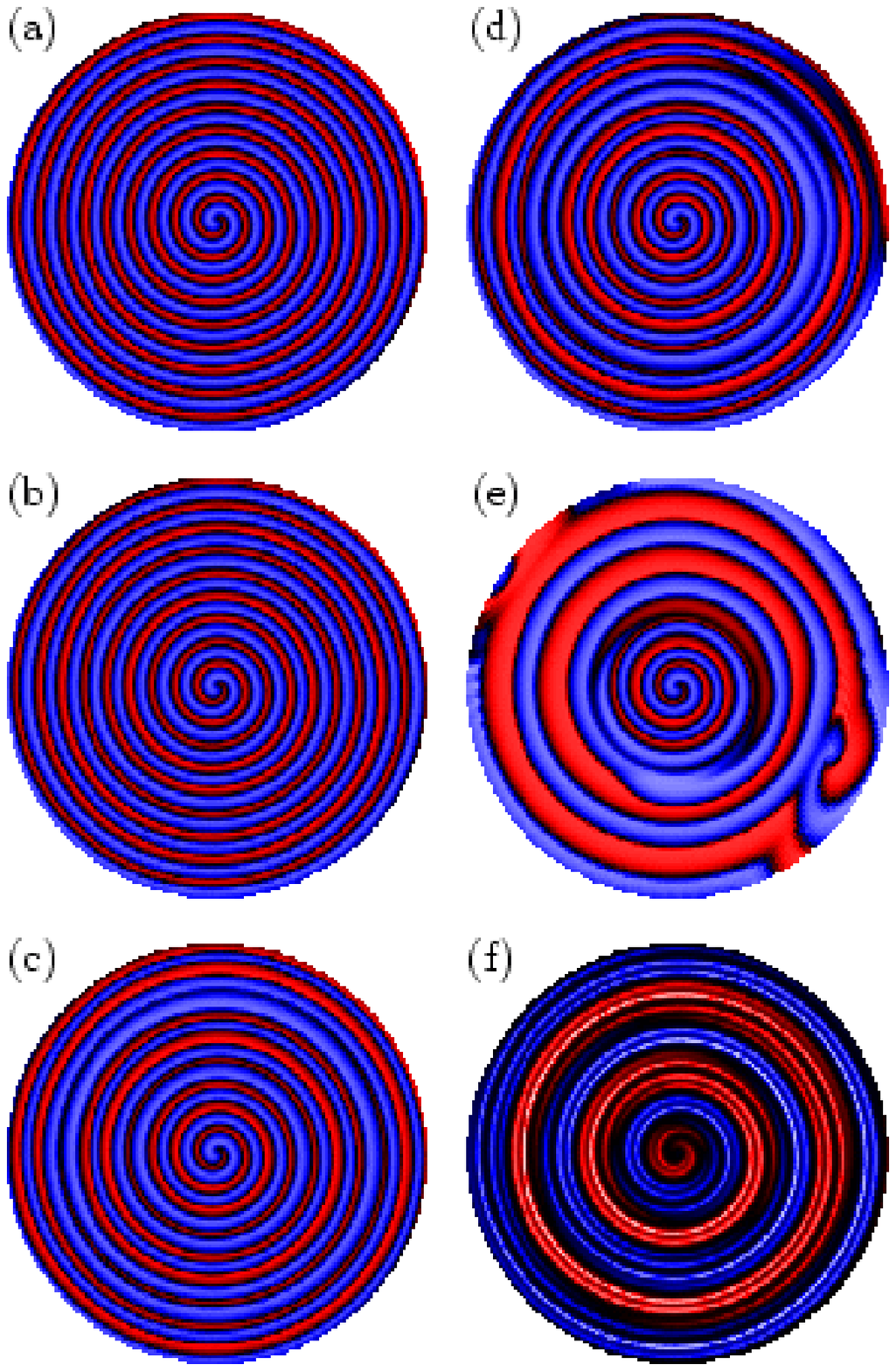}
\caption{Snapshots of evolution from the perturbed steady spiral leading to
  far-field breakup (solid green curve in figure~\ref{fig:far_perturb_ts}).
  The $u$-field is shown.  (a) $t=0$, (b) $t=50$ ($\simeq 12$ periods), (c)
  $t=80$ ($\simeq 19$ periods), (d) $t=85$, (e) t=120 ($\simeq 29$ periods),
  (f) eigenfunction.  Parameters are as in figure~\ref{fig:eval}(b). }
\label{fig:far_perturb_vis}
\end{center}
\end{figure}

Next we briefly consider far-field breakup. We have directly computed the
eigenfunction associated with absolute instability causing far-field breakup,
figure~\ref{fig:s2evec}(a).  The leading eigenfunction shows exactly the long
wavelength modulation of the steady spiral expected for this
instability~\cite{ialalkaw92,mbmo98,lbatmb03,qojf96,smTeK98,lzqo00}.
Figures~\ref{fig:far_perturb_ts} and \ref{fig:far_perturb_vis} show the
dynamics following from the steady spiral perturbed by the
unstable eigenfunction. The dynamics are initially that of exponential growth
with the expected growth rate. The growth becomes nonlinear and
long-wavelength modulation of spiral becomes visible (time 80 in figure
\ref{fig:far_perturb_vis}). Shortly thereafter the spiral breaks near the
domain boundary. At these parameter values the eigenfunction grows weakly with
radius, as seen in figure~\ref{fig:s2evec}(a), and for this reason one would
expect a preference for breakup near the domain boundary.  However, in this
case the qualitative character of eigenfunction depends sensitively on
parameters and for slightly different parameters the eigenfunction may decay
weakly with radius.

The far-field case is similar to the core breakup case in most other respects.
The nonlinear growth in figure~\ref{fig:far_perturb_ts} is faster than linear.
No other eigenvalues appear to play a direct role in the far-field breakup.
Figure~\ref{fig:far_perturb_ts} shows the evolution from an initial condition
formed from the eigenfunction corresponding to the complex-conjugate point
eigenvalues near the imaginary axis [(b) in figure~\ref{fig:s2evec}].

\section{Conclusions}
\label{sec:conclusions}

In this paper we have examined in detail the linear stability spectra and
associated eigenfunctions for spiral waves in large domains.  Everywhere,
except possibly near the edges of the absolute spectra, numerically computed
eigenvalues and eigenfunctions agree extremely well with the results of
Sandstede and Scheel.  Our results answer the question posed at the
outset. Absolute spectra can be relevant and predictive for typical domain
sizes used in studies of spiral waves.  Even in domains containing only a few
spiral wavelengths (the case $R=20$) eigenvalues show signs of the absolute
spectrum - they lie along curves located roughly in the correct part of the
complex plane.  For domains containing five spiral wavelengths or more ($R
\gtrsim 40$) eigenvalues lie quite close to the absolute spectra.  Of course
these results are for the particular equations and parameters studied here and
absolute spectra will not necessarily be such good predictors for domains of
these sizes in other systems.  Nevertheless, in at least two cases, one with
excitable dynamics and one with oscillatory dynamics, absolute spectra are
predictive.

The computed eigenvalues support convergence to the absolute spectrum
inversely with the domain radius, at least away from the edge of the absolute
spectrum.  In most cases eigenfunctions agree with the exponential forms
deduced by Sandstede and Scheel.  This is even the case for point eigenvalues
not associated with the absolute spectrum.  Near the edges of the absolute
spectrum, however, eigenfunctions do not exhibit exponential growth at large
radius, even in the largest domains we have considered. While results suggest
that this is due to finite-size effects, more work is necessary to understand
the behavior of eigenvalues and eigenfunction near the edges of the spectrum.

We have computed the positive point eigenvalues giving rise to both core and
far-field breakup of spiral waves.  The essential difference between the two
cases is the form of the eigenfunctions. For core breakup the eigenfunction
has a maximum not far from the core of the spiral and decays at large
radius. For far-field breakup the eigenfunction grows with radius.  Moreover,
the far-field eigenfunction shows long-wavelength modulation known to precede
far-field breakup.  Nonlinearity also plays a role in breakup and we have
presented nonlinear simulations showing the destabilizing effects of
nonlinearity in each case.

The most intriguing aspect of this work is that all point eigenvalues we have
found appear near edges of the absolute spectra. This may be a coincidence,
but it would not seem so from figure~\ref{fig:eval}.  We leave this as an open
problem.

\section*{Acknowledgments}
We are very grateful to Bjorn Sandstede and Arnd Scheel for helpful
discussions and for kindly providing data used for many of the comparisons
presented.

\bibliographystyle{siam}
\bibliography{spiral}

\end{document}